\documentclass{article}

\usepackage{PRIMEarxiv}

\usepackage[utf8]{inputenc} 
\usepackage[T1]{fontenc}    
\usepackage[hidelinks]{hyperref}       
\usepackage{url}            
\usepackage{booktabs}       
\usepackage{amsfonts}       
\usepackage{nicefrac}       
\usepackage{microtype}      
\usepackage{lipsum}
\usepackage{fancyhdr}       
\usepackage{graphicx}       
\graphicspath{{media/}}     

\usepackage[backend=biber]{biblatex}
\usepackage{amssymb}
\usepackage{tabularx}
\usepackage[acronym]{glossaries}
\glsdisablehyper

\def\paperRunningTitle{A Visual Approach for Health Information Exploration}
\def\paperAbstract{%
The effective and targeted provision of health information to consumers, specifically tailored to their needs and preferences, is indispensable in healthcare. With access to appropriate health information and adequate understanding, consumers are more likely to make informed and healthy decisions, become more proficient in recognizing symptoms, and potentially experience improvements in the prevention or treatment of their medical conditions. 
Most of today’s health information, however, is provided in the form of static documents. 
In this paper, we present a novel and innovative \emph{visual health information system} based on adaptive document visualizations. 
Depending on the user's information needs and preferences, the system can display its content with document visualization techniques at different levels of detail, aggregation, and visual granularity. 
Users can navigate using content organization along sections or automatically computed topics, and choose abstractions from full texts to word clouds. 
Our first contribution is a formative user study which demonstrated that the implemented document visualizations offer several advantages over traditional forms of document exploration. 
Informed from that, we identified a number of crucial aspects for further system development.
Our second contribution is the introduction of an \emph{interaction provenance visualization} which allows users to inspect which content, in which representation, and in which order has been received. 
We show how this allows to analyze different document exploration and navigation patterns, useful for automatic adaptation and recommendation functions. We also define a baseline taxonomy for adapting the document presentations which can, in principle, be leveraged by the observed user patterns. 
The interaction provenance view, furthermore, allows users to reflect on their exploration and inform future usage of the system.}
\def\paperAcknowledgement{This work was funded by the Austrian Science Fund (FWF) as part of the project `Human-Centered Interactive Adaptive Visual Approaches in High-Quality Health Information' (\apluschis; Grant No. FG 11-B).}
\def\keywordOne{Document Visualization}
\def\keywordTwo{Adaptive Visualization}
\def\keywordThree{Interaction Analysis}
\def\keywordFour{Health Information}
\def\keywordFive{Evaluation}

\newacronym{dl}{DL}{Document Library}
\newacronym{toc}{ToC}{Table of Contents}
\newacronym{wc}{WC}{Word Cloud}
\newacronym{hwc}{HWC}{History WC}
\newacronym{topicb}{ToB}{Topicbar}
\newacronym{topiccloud}{ToC}{Topiccloud}
\newacronym{tileb}{TiB}{Tilebar}
\newacronym{snps}{Snps}{Snippets}
\newacronym{fulltext}{FullT}{Fulltext}
\newacronym{is}{IS}{Image Slider}
\newacronym{chis}{CHIS}{Consumer Health Information Systems}
\newacronym{apchis}{A\textsuperscript{+}CHIS}{Adaptive CHIS}
\newacronym{ttwodm}{T2DM}{Type 2 Diabetes Mellitus}
 
\newacronym{cwt}{CWT}{Cognitive Walkthrough}
\def \cwt {\acrshort{cwt}}

\newcommand{\DocumentLibrary}{\acrlong{dl}}
\newcommand{\dl}{\acrshort{dl}}
\newcommand{\TableOfContents}{\acrlong{toc}}
\newcommand{\toc}{\acrshort{toc}}
\newcommand{\WordCloud}{\acrlong{wc}}
\newcommand{\wc}{\acrshort{wc}}
\newcommand{\Topicbar}{\acrlong{topicb}}
\newcommand{\tob}{\acrshort{topicb}}
\newcommand{\Tilebar}{\acrlong{tileb}}
\newcommand{\tib}{\acrshort{tileb}}
\newcommand{\Snippets}{\acrlong{snps}}
\newcommand{\snps}{\acrshort{snps}}
\newcommand{\ImageSlider}{\acrlong{is}}
\newcommand{\is}{\acrshort{is}}
\newcommand{\isL}{\acrshort{is}\textsubscript{l}}
\newcommand{\isS}{\acrshort{is}\textsubscript{s}}

\newcommand{\hwc}{\acrshort{hwc}}
\newcommand{\Searchbar}{Searchbar}
\newcommand{\Startscreen}{Startscreen}
\newcommand{\Fulltext}{\acrlong{fulltext}}
\newcommand{\fullt}{\acrshort{fulltext}}
\def \apluschis {\acrshort{apchis}}
\def \chis {\acrshort{chis}}

\newcommand{\unit}[1]{\ \mathrm{#1}}

\def \taskWcOne {\wc1}
\def \taskWcTwo {\wc2}
\def \taskWcThree {\wc3}
\def \taskWcFour {\wc4}
\def \taskTibOne {\tib1}
\def \taskTibTwo {\tib2}
\def \taskTibThree {\tib3}
\def \taskTibFour {\tib4}
\def \taskIsOne {\is1}
\def \taskIsTwo {\is2}
\def \taskIsThree {\is3}

\def \POne {P01}
\def \PTwo {P02}

\def \PSix {P06}
\def \PSeven {P07}

\def \PTen {P10}

\def \PTwelve {P12}

\newcommand{\figref}[1]{Figure~\ref{#1}}
\newcommand{\secref}[1]{Section~\ref{#1}}
\newcommand{\tableref}[1]{Table~\ref{#1}}
\newcommand{\eg}{e.g.,}
\def\mediumwidth{.6}%
  
\title{A Visual Approach for Health Information Exploration: Adaptive Levels of Visual Granularity and Interaction Analysis
\thanks{\textit{\underline{Citation}}: 
\textbf{Preprint submitted to the Journal of Universal Computer Science}}
}

\author{
    Stefan Lengauer$^1$\thanks{\textit{\underline{Corresponding Author}}: \texttt{s.lengauer@tugraz.at}}, 
    Lin Shao$^2$, 
    Hossein Miri$^3$, 
    Michael Bedek$^4$, 
    Cordula Kupfer$^4$, 
    Maria Zangl$^4$,\\
    \textbf{Bettina Kubicek$^4$,
    Barbara Dienstbier$^5$, 
    Klaus Jeitler$^{5,6}$, 
    Cornelia Krenn$^5$, 
    Thomas Semlitsch$^5$,}\\
    \textbf{Carolin Zipp$^5$, 
    Dietrich Albert$^4$, 
    Andrea Siebenhofer$^{5,7}$, and 
    Tobias Schreck$^1$} \vspace{2mm}\\
  $^1$Institute of Visual Computing, Graz University of Technology \\
  $^2$Fraunhofer Austria Center for Data Driven Design\\
  $^3$International School of Engineering, Chulalongkorn University\\
  $^4$Institute of Psychology, University of Graz\\
  $^5$Institute of General Practice and Evidence-based Health Services Research, Medical University of Graz\\
  $^6$Institute for Medical Informatics, Statistics and Documentation, Medical University of Graz\\
  $^7$Institute for General Practice, Goethe University Frankfurt am Main
}

\begin{document}
\maketitle

\begin{abstract}
\paperAbstract
\end{abstract}

\keywords{\keywordOne \and 
\keywordTwo \and
\keywordThree \and
\keywordFour \and
\keywordFive}

\section{Introduction}

\acrfull{chis} have become an important tool in modern healthcare and serve a variety of functions. 
They can provide consumers with a comprehensive understanding of a disease, specifically addressing general knowledge of health-related issues, effects, ways, and measures to maintain and possibly restore health. 
They can also enable early detection, diagnosis, treatment, palliation, rehabilitation, and follow-up care for diseases, along with associated medical decisions, care, and coping strategies for daily life with these diseases~\cite{RN1}. 
Typically, \chis\ are provided in a static and linear manner, meaning the same medical content is presented to everyone in the same structure. 
However, a linear reading or navigation may not be the best solution for everyone to extract relevant information because patients differ in terms of their prior knowledge, information needs, personal preferences and styles, and health situation, which may depend on factors such as gender, age, personality, and perception~\cite{RN3}. 
To address this problem, an \emph{adaptive and interactive visual \chis\ that supports document exploration with adaptive focus and detail views} is needed. 
By tailoring content to individual user needs and preferences, such as adjusting the level of detail or focusing on specific topics, this type of system would allow users to personalize their experience.

The primary research goal of this work is to develop novel concepts for advanced, interactive, adaptive, and visual \chis\ (called \apluschis). 
We selected \acrfull{ttwodm} as a pilot disease for our research because it is a complex disease with high prevalence and relevance to the public health system. 
Also, its progression over time requires changes and adaptations, including tailored knowledge, flexible treatment, new drug classes, improved patient education, sustainable follow-up practices, and screening for complications. Managing \acrshort{ttwodm} is still a difficult and time-consuming task, as it is common, serious, and under-treated. 
This is a major challenge for healthcare services, and patients and therapists must be prepared to deal with it effectively. 
Consequently, those affected by \acrshort{ttwodm} have an ongoing requirement for timely and pertinent information.~\cite{RN6}.

\begin{figure*}[ht!]
    \centering
    \includegraphics[width=\linewidth]{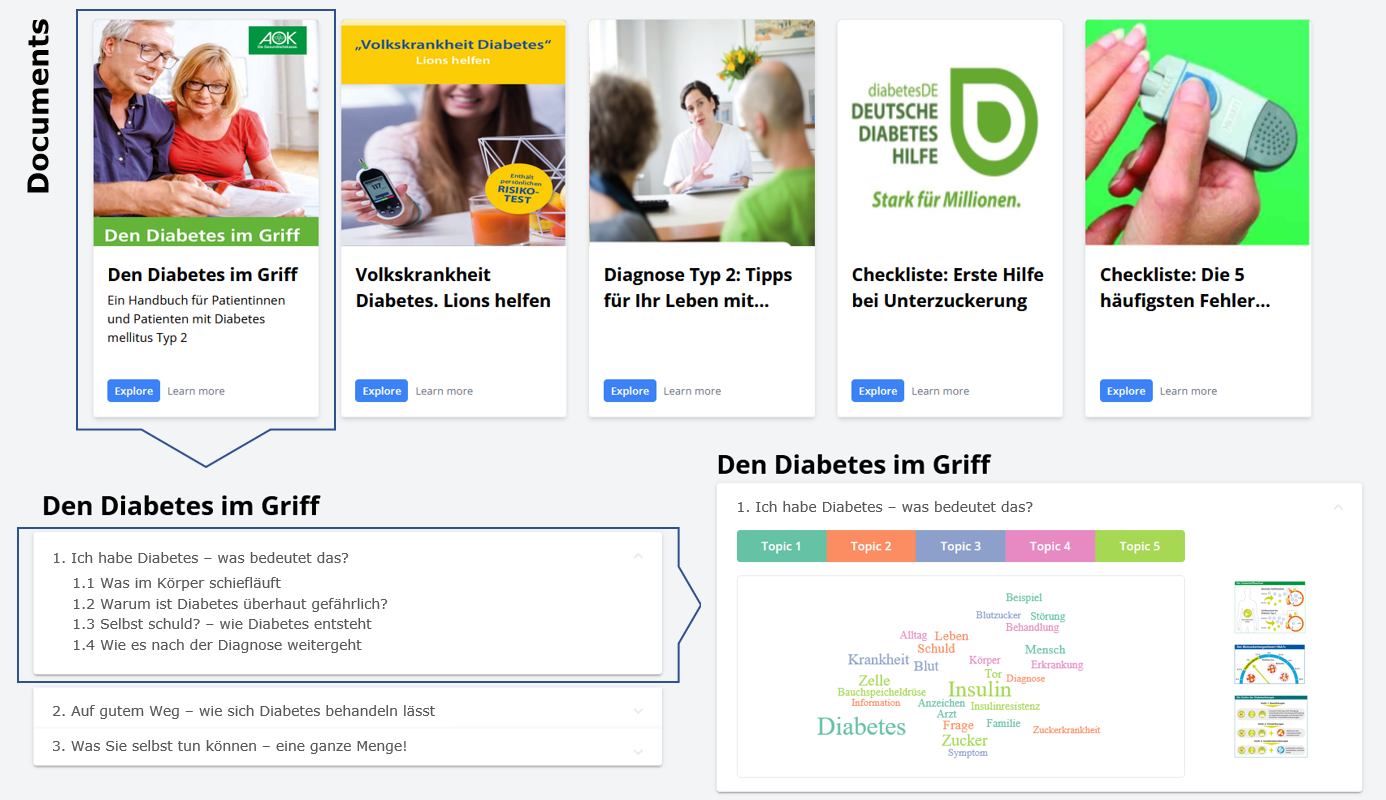}
    \caption{
    The proposed \apluschis\ supports dynamic levels of detail. 
    A \DocumentLibrary\ (top) allows users to select one particular document they want to explore. 
    At document level, an interactive \TableOfContents\ (bottom left) preserves the global linear structure of the document while a chapter's substructure and content is visualized with dedicated visualization techniques for textual and pictorial data.
    }
    \label{fig:concept-dl-and-toc}
\end{figure*}

In this paper, we present a novel visual document exploration system with multi-dimensional adaptivity to help health information consumers better understand medical content by combining close and distant reading approaches.
It is targeted towards non-medical users of all adult age groups, which have either \acrshort{ttwodm}, are a relative to a \acrshort{ttwodm} patient or are interested in the disease for another reason. 
After a development and evaluation phase the \apluschis\ should  eventually be freely discoverable on the Internet. 
We work thus with an unpremeditated userbase in mind, which should be able to use the \acrshort{chis} intuitively, without the need for a supervised introduction. 

To this end, we propose an innovative document exploration system that provides multi-level navigation from high-level (topic overview) to mid-level (keyword occurrence and highlighting) to low-level (full text) views (\figref{fig:concept-dl-and-toc}). 
The basic idea is to allow users to efficiently navigate through documents, overview the content, find topics of interest, and finally switch to a close read on specific information contents. 
In our system, we make use of well-known document visualization approaches to enhance the learning process of medical content. To visualize high-level structures of a document, we propose dynamic table-of-content which represents sub-chapters by means of a \WordCloud\ containing keywords from a topic-modeling approach. High-level structures and mid-level document information are linked by using Tile Bars as visual navigator. 
The visual navigator shows topic occurrences within the underlying document and allows users to quickly explore the content by text snippets.
Although these text visualization techniques are not novel themselves, we tie them together in an integrated implementation prototype and adapt them for the specific requirements of the health domain. The system also serves as a platform to test and evaluate approaches for adaptive document and interaction provenance visualization (cf.\ also Section \ref{sec:interaction-analysis}).
Our system introduces the notions of  level of detail and adaptive visual presentations for document-based health information exploration for \acrshort{ttwodm}. 
Existing \acrshort{chis} are largely static in nature and do not adapt either of these dimensions to their users.

As a consequence, we conducted a user study to characterize the usage behavior of health information seekers adopting our approach. 
We show the usability of our system by comparing linear reading with our multi-level approach, and illustrate the results by two provenance visualizations.

This paper presents a comprehensive extension of one of our previous papers \cite{lin2023ivapp} delving further into our developed system and offering additional insights and analyses. 
Besides a more verbose and detailed outline of our work, this paper comes with the following additions with regards to its predecessor:
\begin{enumerate}
    \item An extension to our system design with an updated visual representation as well as newly-added components such as a \DocumentLibrary\ for the exploration at documents level and an alternative to the Word Cloud representation (\acrlong{topiccloud});
    \item A through analysis of the interaction data obtained through supervised evaluations with actual users;
    \item Different customized visual analytics tools for analysing and evaluating said interaction data.
\end{enumerate}
In the next section (\secref{sec:related-work}) we provide an overview on relevant previous works from the consumer health information domain and reveal the research gap regarding adaptive systems for said domain, before our proposed system design is outlined in \secref{sec:proposed-design}. 
\secref{sec:evaluation} describes a formative evaluation that incorporates quantitative and qualitative methods, performed by a number of participants on an implementation prototype. 
Following this formative evaluation it was observed that the rich interaction data collected in its course merits the effort of a more detailed analysis and the development of additional visual analytics tool (\secref{sec:interaction-analysis}).
On that note, we also investigated the suitability of user interactions for proposing adaptive visualizations. 
\secref{sec:discussion} and \ref{sec:conclusion} conclude the paper with a thorough discussion regarding next steps and open research challenges. 

\section{Related Work} \label{sec:related-work}

In this section, we present an overview of important visualization techniques that have inspired our approach and review the need for adaptive and interactive consumer health information systems.

\subsection{Visualization Techniques for Text Documents and Health Care}
Visualizing large text corpora is a challenging task. 
Usually, the involved data sets are inherently complex, containing structural and content-related information. 
Most linguistic and text visualization approaches rely on text-mining techniques to reveal semantic information from raw text data. 
Therefore, simple statistical processing (\eg\ word frequency and bag-of-words concept) as well as natural language processing approaches (\eg\ named-entity recognition, relationship extraction and sentiment analysis) may be used~\cite{10.1002/widm.1071, 7156366}.

A widely-used visualization technique for text data is the Word Cloud representation (also known as Tag Cloud) which presents an overview of the most frequent or important words by using different type or font sizes~\cite{6758829}.
This technique is also known as distant-reading technique~\cite{moretti2005graphs} and allows users to approach literature in a new way.
Instead of reading texts in the traditional way, i.e., linear or close-reading, the focus of distant-reading approaches is to count, graph, and map textual data by a visual representation~\cite{janicke2015close}.
In recent years, much research has been conducted on distant-reading and Word Cloud visualizations. 
For instance, Kim et al.~\cite{5718617} proposed WordBridge, which utilizes graph-based visualization techniques to connect multiple Word Clouds with information-rich edges. 
Further extensions of Word Cloud exist that focus on semantic contour lines~\cite{2011.01923.x} and images~\cite{doi:10.1177}. 
In our work, we rely on traditional Word Clouds to foster distant-reading within single documents.

For the exploration of larger document collections, additional document features such as metadata information and co-authorships, could be considered to gain a better understanding of the contents of those documents~\cite{6392833, 7583708}. 
Another interesting approach by Strobelt et al.~\cite{5290723} called Document Cards, utilizes a mixture of images and important keywords to visualize key semantics of a document. 
To visualize distributional properties within a document, Tile Bars~\cite{hearst1995tilebars,keim2007literature} could be considered, which is a compact pixel-based visualization technique that reveals the relative length of a document and the relative frequency of one or more query terms. 
In our work, we utilize Tile Bars to represent the relative frequency and distribution of terms from a Word Cloud.

Data visualizations are becoming increasingly important for various fields of application, as well as in healthcare. 
Visual representations may help patients as well as physicians to gain a better understanding of health records, \eg\ information on medical diagnostics, treatments, and health histories~\cite{HCI-039}. 
For example, the LifeLine system was among the first exploration systems that supports visual patient treatment histories~\cite{10.1145/286498.286513}. 
An extensive survey about visualization techniques for electronic health records and population health records are given in~\cite{DBLP:journals/cgf/WangL22}.

Recently, many of the mentioned document visualization techniques are also applied in a medical context. 
For instance, Facetatlas by Cao et al.~\cite{5613456} used linked Word Clouds to visualize entity-relational text document of diseases such as Type 1 and Type 2 Diabetes
Mellitus. 
The linked Word Clouds are used to represent global relations by using a density map and local relations by using edge bundling techniques. 
Another interesting multifaceted text visualization is SolarMap~\cite{6137214} which combines a labeled contour-based cluster visualization with a radially-oriented word cloud.
Furthermore, SolarMap can visualize topic distribution of entities from one facet together with keyword distributions that convey the semantic definition along a secondary facet.

With the advent of novel visualization techniques in different domains, visualization literacy, i.e., user understanding and discovery of visual patterns, is becoming increasingly important. 
Developing visual literacy is essential to support cognition and evolve toward a more informed society~\cite{doi:10.1177/14738716221081831}. 
In our work, we intend to increase visual and health literacy by gathering user information during exploration and providing adapted health information based on that.

\subsection{Need for Adaptive \chis} \label{subsec:chis}

As part of this work, we examined current sources of \chis\ related to \acrshort{ttwodm} across multiple media platforms, including websites, digital documents (PDFs), print media, apps, and videos. Our goal was to identify elements and modes of presentation within a representative sample of these sources that users can customize to their needs and preferences.
Our results suggest that the potential for adaptation in existing CHIS is only realized to a limited extent.
We did not find any adaptive elements in print media or digital documents (PDF) while websites, apps, and videos offer some customization options related to presentation format, such as adjusting font size and color. 
Some \chis\ also included features such as text-to-speech or language-switching~\cite{RN7, RN8}.
However, in terms of personalized medical information, only a few \chis\ had mechanisms to pre-filter medical content based on a user's diabetes profile~\cite{RN9}.
Most \chis\ included a standard table of contents, with or without hyperlinks to the respective chapters. Some sources also contained links within the text or cross-references to other sections or chapters. 
However, none of the \acrshort{ttwodm} \chis\ we analyzed used a visual document exploration system with multi-dimensional adaptivity for health information consumers.

These results show that existing \chis\ on \acrshort{ttwodm} fall short of the potential of presenting health information in an \emph{interactive}, \emph{adaptive} and/or \emph{personalized} way, while there is evidently a need for it. 
The knowledge domain of \acrshort{ttwodm} is complex and comprehensive, with a wide range of information sources (brochures, websites, medical doctors, etc.) and high diversity of topics (such as symptoms, treatments, nutrition, etc.). 
This might be overwhelming for laypersons without medical expertise seeking knowledge in the field. Such complex information situations usually put a high \emph{intrinsic cognitive load}~\cite{sweller2005implications} on the working memory during information processing and often lead to information seekers applying heuristics and cognitive biases at every stage of information processing. 
Such cognitive biases, misconceptions, and even myths about \acrshort{ttwodm} may lead to unhealthy behavior with severe health-related consequences.

An \emph{interactive} \chis\ has the potential to (i) track behavioral patterns and explicit feedback of consumers, (ii) interpret these indicators in terms of certain cognitive biases (\eg\ the confirmation bias), and (iii) intervene if necessary (\eg\ by suggesting other pieces or sources of information). 

An \emph{adaptive} \chis\ can match the information units to the users and their current information needs. 
It can thus balance the \emph{intrinsic cognitive load} to a medium level and ensure that the consumer is neither too bored nor too overwhelmed. 
This is in line with the transfer of Vygotsky’s concept of the \emph{zone of proximal development} to digital learning environments~\cite{luckin2010re}, the outer fringe as suggested by the competence-based knowledge space theory~\cite{heller2006competence} and constitutes a solid basis for an enjoyable flow for consumers~\cite{schiefele2011skills}. 
All these theories emphasize that a medium difficulty of information units lead to a successful processing outcome. The intermediate goal is to successively reach more advanced levels of learning outcomes as suggested by Krathwohl~\cite{krathwohl2002revision} which means not simply remembering information, but also applying and evaluating it.

A \emph{personalized} \chis\ can foster a consumers’ personal commitment to engage with the system and information, and thus help to close the `intention-behaviour gap’ or `attitude-action gap' \cite{schwarzer2008modeling}. 
This is the ultimate goal of any \chis. 
We strive to achieve this through different added values of our advanced \chis\ compared to more `traditional’ digital \chis\ (\eg, a brochure in PDF format or plain webpage): the guarantee of high quality and evidence-based medical information, the reduction of complexity to a medium level, and the recommendation of information units that fit a consumers’ information needs. 
In addition, tools and functionalities to get an overview of the knowledge domain, to efficiently answer certain questions, and to easily navigate through different sub-topics will be offered for good user experience. With our formative evaluation activities (\secref{sec:evaluation}) we monitor progress towards these goals.

\section{Adaptive Document Exploration Design} \label{sec:proposed-design}
In the following \secref{sec:design-requirements} we discuss the design considerations, taking into account the intended userbase and use case scenarios. 
Informed by this, the resulting system design is outlined in detail in \secref{sec:implementation}.

\subsection{Design Requirements/Considerations}\label{sec:design-requirements}
In view of our intended users, who consists of all adult age groups, genders, and digital proficiency we decided to rely on document visualization techniques that are intuitive to understand and easy to use. 
Although the participants of the formative evaluation study which was conducted within the scope of this project (\secref{sec:evaluation}) received a brief introduction into visual components of the system, we want to support the scenario that an uninformed user, which discovers the \acrshort{chis} on the Internet is able to use it without any prior supervised introduction. 
That is, visualizations should base upon tried and tested concepts, which have been shown to be usable by the vast majority of people. 
Yet, at the same time, more advanced users should have to option to alternatively work with more complex visualizations, which allow for a more efficient exploration at the cost of a steeper learning curve.

\begin{figure*}[ht!]
    \centering
    \includegraphics[width=\textwidth]{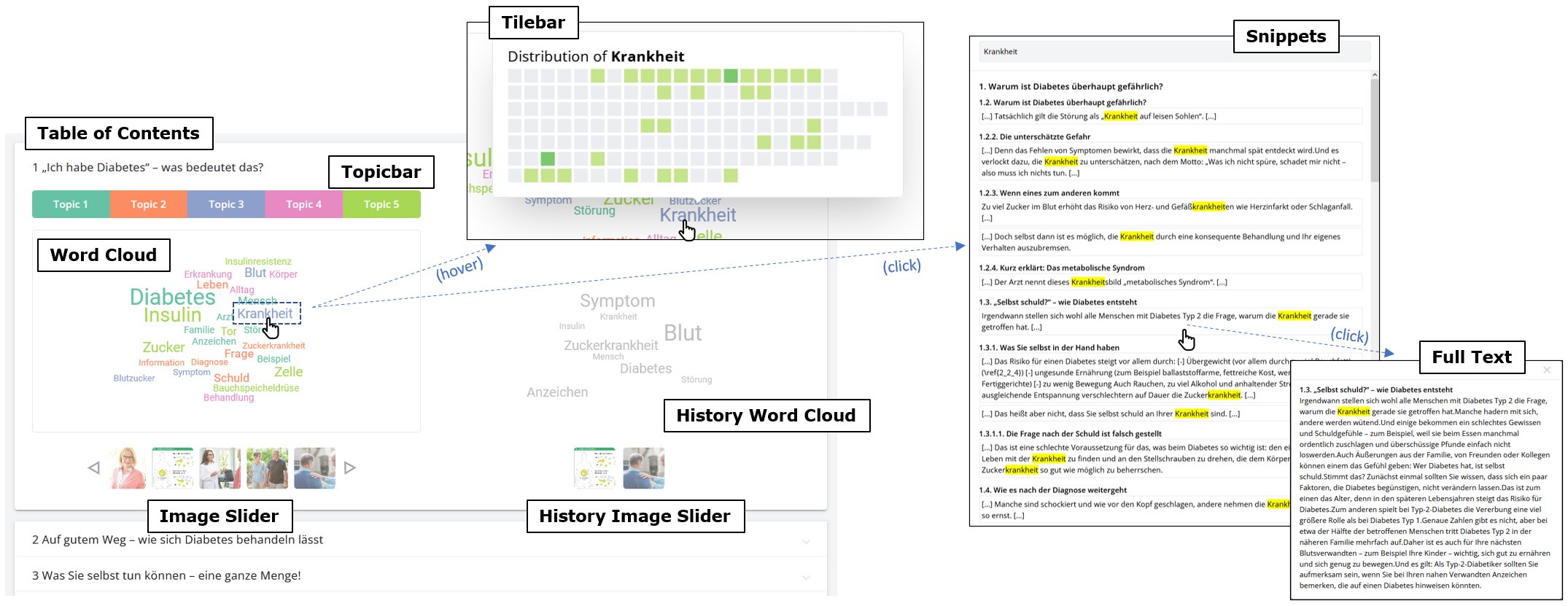}
    \caption{
    The main components of \apluschis\ shown by an example of exploring a German diabetes health brochure \cite{aok}:
    \acrfull{toc}, \acrfull{wc}/\acrfull{hwc}, \acrfull{is}, \acrfull{tileb}, \acrfull{topicb}, \acrfull{snps}, and \acrfull{fulltext}.
    Different actions (illustrated as blue arrows) allow a user to navigate from one view to another.}    
    \label{fig:exploration-mockup}
\end{figure*}

\subsection{Implementation}\label{sec:implementation}
Informed by the considerations above we designed a system comprised of several components, which support the exploration of documents at different levels of visual granularity. 
Starting from a \acrfull{dl} showing an overview of the documents supported by the \apluschis, a user is able to explore individual documents on both high and low levels (\figref{fig:exploration-mockup}).
For the prior, the system provides an expandable and interactive \acrfull{toc} while the latter is enabled though a series of text abstraction methods such as \acrfull{wc}, fingerprinting in the form of a \acrfull{tileb}, and topic modeling by means of a \acrfull{topicb}.
Pictorial content is provided in an \acrfull{is} component.
On the lowest level, a user can also review sections of the original text sources in the form of \acrfull{snps} or even the untampered full text. 
These different concepts are implemented in different subsystems such that an unintermitted exploration process is possible while alternating between levels.
In the background, we track user interactions to determine which parts of the content have already been visited and consumed by the user. 
This information is also displayed to the user in order to indicate which information has not yet been scrutinized.

\paragraph*{\DocumentLibrary}
The \dl\ is the first view a user is greeted with upon logging into the \apluschis.
All available documents are presented in a grid arrangement with their respective covers and titles (\figref{fig:concept-dl-and-toc}).
Upon hovering over a certain document a preview appears, showing both related metadata and a histogram of the document's most frequent terms. 
This initial view should serve the users in determining which of the documents is the most appropriate for addressing their respective information need.
Clicking on a document transfers the user to its \toc.

\paragraph*{\TableOfContents}
The exploration process on document level is supported by an interactive \toc.
We base this view on a document's inherent linear structure with chapters, sub-chapters, and so on. 
While we aim to preserve the outermost structure (`chapters' in most cases) we abstract all lower levels of structure and content with dedicated text- and multimedia visualizations (\figref{fig:concept-dl-and-toc}). 
To this end, we employ \wc s -- and most recently also an alternative representation (\secref{sec:future-work}) -- for the textual content, \is s for the pictorial content, and further subsystems (\tib\ and \snps) for structure and lower levels of visual granularity.
These dedicated visualizations are interweaved into the linear chapter structure, based loosely on the \emph{document card} design concept by Strobelt et al.~\cite{5290723} -- i.e., different visualization techniques are used to display the textual and visual contents.
On a per-chapter basis, these visualizations can be expanded or collapsed. 

Additionally, already `consumed' content is tracked and indicated with a `history' version of the respective \wc s and \is s. 
Specifically, terms and images are added to these components after they have been reviewed (i.e., clicked on) by the user. 
This \acrfull{hwc} visualizes the context of the exploration to the user. 
Alternatively, it can   be used to display non-clicked terms as to suggest content to the user.

\paragraph*{\WordCloud\ with \Topicbar}
To generate the word clouds, natural language processing is used to extract `significant' terms from each chapter. 
This pre-processing comprises steps such as tokenization (separation/segmentation into individual parts) and stop-word removal (filtering of irrelevant/insignificant words).
Subsequently, the set of remaining words are subjected to a lemmatization (transformation into their canonical form or dictionary form) and grammatically tagged (part-of-speech tagging). 
However, we do not use the resulting normalized words directly to fill the \wc, but instead subject them to the Latent Dirichlet Allocation~\cite{blei2003latent} in order to obtain meaningful topic models on them. 
The appropriate number of topic models per-chapter (5) was determined empirically by visually evaluating the resulting topics for different values in the low integer range.
That is, each topic model is defined by a weighted vector containing a subset of the chapter's extracted nouns. 

The concated vectors of all topics serve as the input for a chapter's \wc, where the weights are used to determine a word's size within the cloud.
For the arrangement of terms, we rely on the \emph{Wordle word cloud} algorithm \cite{steele2010beautiful} while the mapping between words and their belonging topic is established through a qualitative color mapping (\figref{fig:exploration-mockup}, \wc). 
Note that this means certain terms could appear redundantly as topics can exhibit overlapping term compositions.

Since the concurrent display of all topics can be overwhelming, we provide a means to toggle the visibility of individual topics. 
This subsystem -- the \tob\ directly above the \wc\ -- consists of 5 colorized toggle buttons mapping to the respective topic models.
Hovering over a term toggles the \Tilebar\ component for the respective term above it, while clicking it initiates the \snps\ view. 
To this end, we track the interactions -- how often the user has clicked a certain term -- in the \wc.
These click counts are the basis for the so-called \acrfull{hwc} on the right-hand side of a chapter visualization (\figref{fig:exploration-mockup}, \hwc) where the count determines the size of a term in the cloud.
The same hover-and-click interactions as with the `regular' word cloud are possible with the \hwc.

\paragraph*{\Tilebar}

Hovering over a term in the \wc\ triggers the display of the \tib\ component above it, which allows the user to efficiently grasp the term's occurrences over the whole document. 
This visualization is inspired by the \emph{literature fingerprinting} concept by Keim and Oelke~\cite{keim2007literature} which shows various document properties in a drilled-down manner. 
To this end, we compute the respective term's frequency over equal-sized text chunks and visualize the resulting 'intensities' in a colorized fashion, following the linear document structure (i.e., from top to bottom and left to right, see \figref{fig:exploration-mockup}, \tib). 
That is, the rows of the grid symbolize the document's chapters and the columns the ordered text chunks within a chapter.
Cells which stand for text chunks in which the term does not appear at all are filled with uniform gray color.
The \tib\ allows a user to quickly answer such questions as \emph{``does another chapter also cover this topic?''} or \emph{``how frequently is it mentioned overall?''}.

\paragraph*{\Snippets /\Fulltext}
The above mentioned abstractions are vital to gain an overview of the information covered by a document and determine which sections are the most appropriate to answer a specific information need. 
Ultimately however, it is necessary to provide a user with text chunks to enable them to answer their specific information need. 
To address this issue, we added two additional levels of visual granularity.
Firstly, a \emph{\Snippets} view which pops up to the right hand-side of \toc\ if a term in the \wc\ or \hwc\ is clicked.
Within this view, all sentences containing the clicked term are displayed with a highlighting of the term (\figref{fig:exploration-mockup}, \snps). 
Handles at the beginning and the end of a sentence allow to reveal the preceding and succeeding sentence. 
Those can be clicked iteratively to  display larger parts of the document before and after the found position. Alternatively, the section headers, which are also shown in the snippets view, can be clicked to display a section's whole content immediately (\figref{fig:exploration-mockup}, \fullt).
The top of the \snps\ view also contains a \Searchbar, which allows to readily change the term in question.

\paragraph*{\ImageSlider}
Besides the abstractions for textual content (\wc/\hwc), we would like to indicate the presence of a document's pictorial content. 
To this end, we employed an off-the-shelf \is\ component, next to the \wc\ to display a chapter's images (\figref{fig:exploration-mockup}, \is). 
The thumbnails of the images can be viewed directly in this slider, yet if further details are needed, an image can be clicked which shows it together with its caption in a full screen modal dialog (henceforth referred to as \isS\ and \isL\ respectively). 
As this component shows just five images at a time, we aim to determine an image's relevance in order to sort the list of images, resulting in the most relevant images being shown initially.
To determine said relevance, we make two assumptions. 
Firstly, we assume that images without a caption (\eg\ scenic backgrounds at the beginning of a chapter) are rather unimportant. 
Secondly, we split the set of images with captions into two tiers with the first tier being comprised of images showing tables, diagrams, and flow charts, or convey any sort of structured information, while all the others belong to the second tier.
The information whether or not an image has a caption is obtained in the extraction process. 
In a similar fashion to the provenance version of the \wc\ -- the \hwc\ -- we provide an additional image slider in the bottom of the \hwc, showing exclusively the chapter's images which have already been clicked (and thus `consumed') by the user.

\section{Formative Evaluation}\label{sec:evaluation}

The formative study aimed to 
(a) explore how the design of the system and its components would be perceived by users, 
(b) evaluate the system in comparison to a linear and static \acrshort{chis}, 
(c) highlight potential areas for improvement, and 
(d) identify prospective research questions.
To maintain methodological consistency and participant comparability, we focused on one particular document for the evaluation setup. 
That is, we used a \acrshort{ttwodm} information brochure of the German health insurance provider AOK~\cite{aok} as data basis for evaluation. 
The text document in PDF format comprises more than $130$ pages of comprehensive and detailed health information, including figures, tables, and info-graphics. 
Full texts with health information were extracted with Adobe PDFBox\footnote{\url{https://pdfbox.apache.org/}} library and the images were extracted manually. 
Subsequently, the sub-chapters and images were sorted by the main chapters.

\subsection{Participants}
Overall, 12 participants (four females) took part in the study, representing the different potential users of \apluschis\ with ages between 26 and 62 years ($M = 40 \unit{yrs.}$, $SD = 14 \unit{yrs.}$). 
Additionally, the participants had different levels of knowledge and competence. On a 5-point rating scale (from $0 := $ very low to $4 := $ very high) they self-assessed their prior knowledge of \acrshort{ttwodm} ($M = 1.00$, $SD = 1.21$), computer and software skills ($M = 2.25$, $SD = .97$), as well as previous experiences with visualizations ($M = 2.58$, $SD = 1.24$).

\subsection{Procedure} \label{sec:procedure}
 The overall procedure for the participants can be divided into the following phases: 
 (i) \emph{Instruction}, 
 (ii) \emph{Cognitive Walkthrough}, 
 (iii) \emph{Forced Choice} and finally, 
 (iv) \emph{Semi-Structured Interviews.} 
 The phases (ii) to (iv) are described in more detail in according subsections below. 
 The sessions that lasted between 60 and 90 minutes per participant were carried out individually, lead by one investigator. 
 Three of the authors took the role of an investigator, each for four participants. 
 In phase (i), i.e. \emph{Instruction}, participants received a short explanation of basic \chis\ functions, such as search functions, to ensure that they all began the evaluation process from a common usage knowledge base. 
 The participants were then set in a real-world usage scenario: they were asked to imagine that they themselves or someone in their family was diagnosed with \acrshort{ttwodm} during a health check-up. 
 That is why they want to find out more about this disease.

The above-mentioned AOK brochure~\cite{aok} was presented in \apluschis, where participants started the first task with the \toc\ view of the brochure, as well as in a PDF viewer (Adobe Acrobat Reader). 
The comparison with the PDF viewer was chosen not only because this is the original, static format of the brochure, but also because PDF viewers are widely used and therefore people are usually accustomed to them. 
The PDF viewer is therefore a challenging benchmark that allowed us to evaluate the expected added value of \apluschis\ compared to conventional \chis. 
The investigators recorded all audio and on-screen activities and additionally made manual notes of their observations.

\paragraph*{\acrlong{cwt}}
To encourage interaction with \apluschis, showing how intuitive its functions are and how quickly the content can be grasped, participants were given pre-defined tasks to explore \apluschis. 
This evaluation method is known as a \acrfull{cwt}~\cite{hollingsed2007usability}.  
The pre-defined tasks also enabled comparable conditions across all participants for the subsequent evaluation steps. 
We defined them in such a way that they represent realistic search and evaluation tasks in the course of an information search. 
For the purpose of the study, the tasks also had to be linked to a measurable goal achievement, for example, the participants had to find and report a specific piece of information. 
It had to be possible to achieve this goal in both PDF viewer and \apluschis. 
For each tool, approximately the same number of tasks were designed, including tasks from the two categories of `generating an overview' vs. `finding specific information'. 
Initially, we defined 11 tasks: four were aimed at using the \WordCloud\ including \Topicbar\ (\taskWcOne--4),
 four at using the \Tilebar\ (\taskTibOne--4), and three at using the \ImageSlider\ (\taskIsOne--3). 
Since we intended to compare the PDF viewer with \apluschis, two parallel versions were created for the 11 tasks that were comparable in terms of difficulty and functions used. 
For example, the task \taskTibFour\ `In which chapter would you most likely start if you wanted to find out more about blood pressure?' had the parallel version `In which chapter would you most likely start if you wanted to find out more about insulin?'. 
This resulted in a total of 22 tasks (\tableref{tab:CWT}) that each participant processed in a within-subject design via the PDF viewer and \apluschis. 
We balanced task order in terms of system and components to avoid sequential effects. 
In addition to the \cwt, participants were asked to express their thoughts during the tasks (i.e., \emph{think-aloud}). 
The investigators also noted behavioral observations during the \cwt\ tasks to complement on-screen and think-aloud activities.

\bgroup
\newcommand{\theader}[1]{\multicolumn{1}{c}{\textbf{#1}}}
\begin{table*}[ht!]
    \centering
    \caption{The \acrshort{cwt} Tasks.}\label{tab:CWT}
    \begin{tabularx}{\textwidth}{l X X}
        \hline
        \theader{Task} & \theader{Parallel Version 1} & \theader{Parallel Version 2} \\
        \hline\hline
        \taskWcOne & Which contents/subject areas do you think are included in chapter 5? & Which contents/subject areas do you think are included in chapter 3?  \\
        \taskWcTwo & What is the   waist size that creates a greatly increased risk for men? & What daily amounts of beer/alcohol are just acceptable for men? \\
        \taskWcThree & How does type I diabetes mellitus develop? & How does type II diabetes mellitus develop?  
        \\
        \taskWcFour & What contents/terms have you searched for so far? & What contents/terms have you searched for so far?  
        \\
        \taskTibOne & How often does the term `smoking' appear in chapter 1? & How often does the term `stress' appear in chapter 1?  
        \\
        \taskTibTwo & Is it worth reading beyond chapter 2 if you want to learn exclusively about `diastole'? & Is it worth reading beyond chapter 4 if you want to learn exclusively about `medication'?  
        \\
        \taskTibThree & Does chapter 2 give a better insight into the topic of `care' than chapter 6? & Does chapter 6 give a better insight into the topic of `(diabetic) foot' or `foot syndrome' than chapter 3?
        \\
        \taskTibFour & In which chapter would you most likely start if you wanted to find out more about `blood pressure'? & In which chapter would you most likely start if you wanted to find out more about `insulin'? 
        \\
        \taskIsOne & According to an illustration in chapter 5, is a sugar value of 100 mg/dl alarming or safe? & Which 3 stages of diabetes therapy are shown graphically in chapter 4? 
        \\
        \taskIsTwo & Which picture in chapter 6 do you think conveys the most relevant information about type II diabetes mellitus? & Which picture in chapter 6 do you think conveys the least relevant information about type II diabetes mellitus?  
        \\
        \taskIsThree & Search for a graphic on the topic of 'sugar metabolism'. & Search for a graphic on the topic of 'nutrition pyramid'.  
        \\
        \hline
    \end{tabularx}
\end{table*}
\egroup

\paragraph*{Forced Choice} 
Following the \cwt, participants were asked to choose between the PDF viewer and \apluschis\ regarding performance goals of system use. 
This evaluation method is known as \emph{forced choice}. 
The following nine performance goals were evaluated: `Would you rather use Adobe Acrobat Reader or the \apluschis\ to 
(a) get an overview of the domain, 
(b) develop a general understanding on \acrshort{ttwodm}, 
(c) search for specific keywords, 
(d) capture the main content, 
(e) search for specific images, 
(f) get an overview of the most informative images, 
(g) efficiently navigate through different topics of the content, 
(h) get answers to questions you might have in mind, and finally, 
(i) trace past searches'. 
We decided to choose a \emph{forced choice} rather than a questionnaire format with Likert scales for two main reasons: first, as a formative and explorative study we aimed for an explicit comparison between \apluschis\ and a challenging benchmark with regards to the nine above mentioned performance goals. 
Deciding between two alternatives might be easier in particular for the inexperienced participants of our sample. 
Second, considering the sample size which seems reasonable for a first formative evaluation but rather small for a summative one, means and standard deviations which could be derived by Likert scales would not allow for inferential statistical methods.

\paragraph*{Semi-Structured Interviews} 
Lastly, we conducted \emph{semi-structured interviews} with participants to ask open and generic as well as closed and specific questions about \apluschis\ as well as further inquiries on the comparison between the two systems. 
The open questions included for example: `Which system would you rather recommend to a person as a first source of information to get started?' or about the non-linear content exploration in \apluschis\ such as ‘How much does this interactive system encourage you to explore further content?’. 
More specific and closed (yes-no) questions were if the participants had already seen or used the different components of \apluschis\ (such as the \WordCloud\ or \ImageSlider, etc.) prior to the session, if they considered them as helpful and if they would like to use them again.

\subsection{Results and Discussion}
In the following section, we outline and discuss the results, starting from a more global evaluation and continuing with more specific results with respect to the components of \apluschis.

\begin{figure}
    \centering
    \includegraphics[width=\mediumwidth\linewidth]{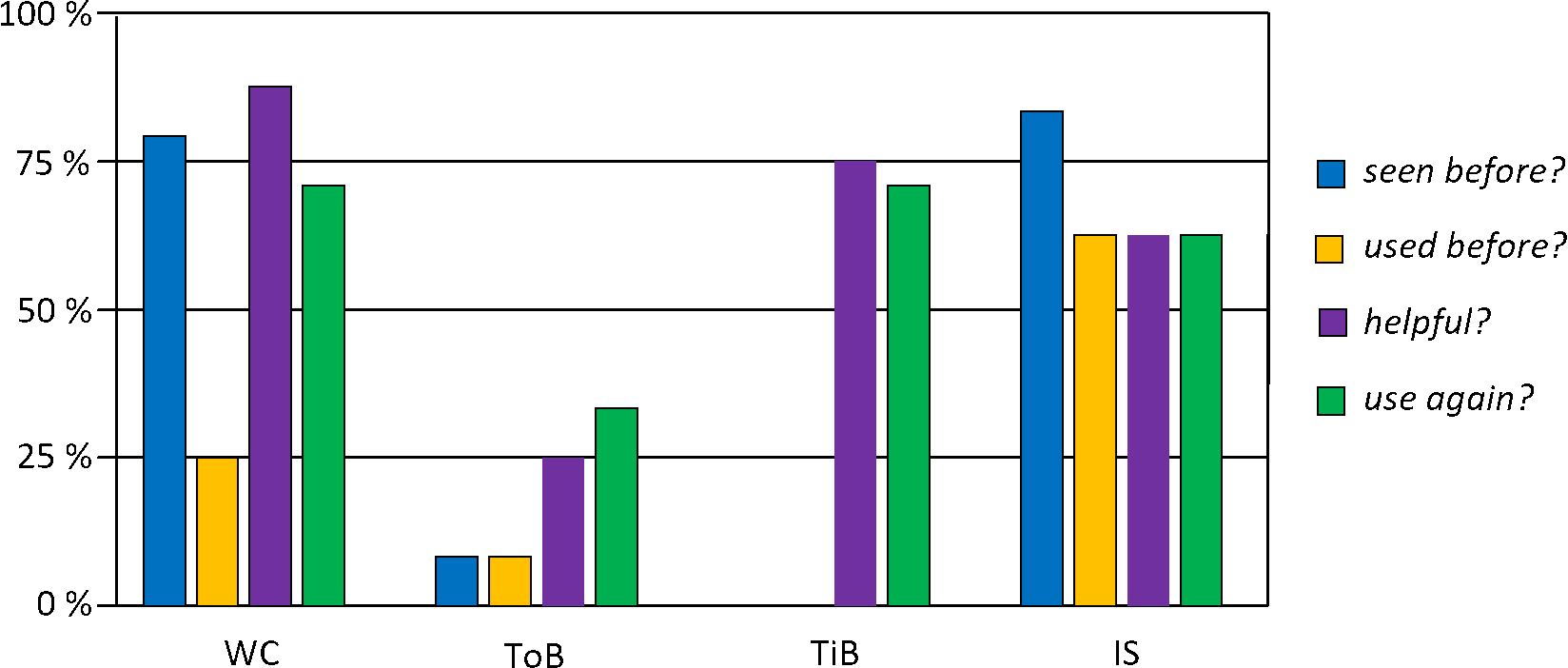}
    \caption{
    The evaluation regarding the prior knowledge and usefulness of \apluschis\ components by the \cwt\ participants.
    }
    \label{fig:seen_used_hepful_useagain}
\end{figure}

\paragraph*{Global Evaluation} 
As outlined in the previous section, in the course of the semi-structured interviews, participants were asked closed questions if they had already seen or used a 
(i) Word Cloud, 
(ii) Topic Bar, 
(iii) Tile Bar, or 
(iv) Image Slider prior to the session. 
In addition, they were asked if the components were considered as helpful and if they would like to use them again. 
A `yes-answer' has been coded as `1', a `no-answer' as `0', and indifference as `0.5'.
\figref{fig:seen_used_hepful_useagain} shows the results of this global evaluation: while \WordCloud\ and \ImageSlider\ were well known by participants, as around 75\% had seen it before, no participant had come across a \Tilebar\ and only one has come across a \Topicbar\ prior to the session. 
The \ImageSlider\ had been used most often before (ca. 63\%). 
Interestingly, the \WordCloud\ had rarely been used despite being well known. 
Several participants noted that they were familiar with a \WordCloud\ as an overview graphic, but not as an interactive element. 
Regarding helpfulness and future use, more than half of the participants found the \WordCloud, the \Tilebar, and the \ImageSlider\ helpful and would consider using them again. 
However, the \Topicbar\ in its current form was not perceived as very helpful, and only 25\% would consider using it again.

\paragraph*{Evaluation of \apluschis\ Components} 
The \cwt\ tasks provided information on how the participants used the different components. 
We observed mixed results regarding the intuitiveness of searching with the \WordCloud. 
Some participants were not hesitant to use the \WordCloud\ and performed efficient searches for keywords and contents, as one participant explained: \emph{‘I simply looked at the largest terms’ (\PTwo)}. 
Others preferred the use of more familiar features, such as the \TableOfContents\ or the \Searchbar. 
However, the \WordCloud\ seemed to be a tool that was easy to learn. We observed that participants often started to use the \WordCloud\ when other features were not perceived as helpful due to the nature of the task or because the traditional search was simply inefficient.

Similarly, we observed that the \Tilebar\ component was mostly intuitive and easy to learn. Several participants were completely unfamiliar with the \Tilebar\, but most were able to quickly understand its function and extract information from it, as this quote demonstrates: \emph{‘There is a box [tile] in chapter 2 and then no more boxes in the chapters after that, so it [the searched term] is not mentioned anymore’ (\PSix)}. 
Overall, the use of \Tilebar\ allowed the participants to efficiently search for the desired information. 
With regard to the design, the participants identified a need for optimization; for example, more contrasting colors for the tiles and a clearer labeling of the chapters within the \Tilebar\ would promote more intuitive use.

Finally, the \ImageSlider\ was perceived as rather intuitive and efficient for finding images because a number of participants already knew it from other applications and the operation was thus familiar. 
However, using the search bar was often preferred over browsing the \ImageSlider, particularly because the latter was time-consuming in chapters with many images, as this participant explained: \emph{‘I don't want to click through all of them; they are unnecessary and very small. 
They don't contain any information; they are just photos without text’ (\PSeven)}. 
This quote also highlights the desired reduction of unnecessary images as well as the wish for more context for those images that were deemed not self-explanatory, as expressed by several participants.

\begin{figure}[ht!]
    \centering
    \includegraphics[width=\mediumwidth\linewidth]{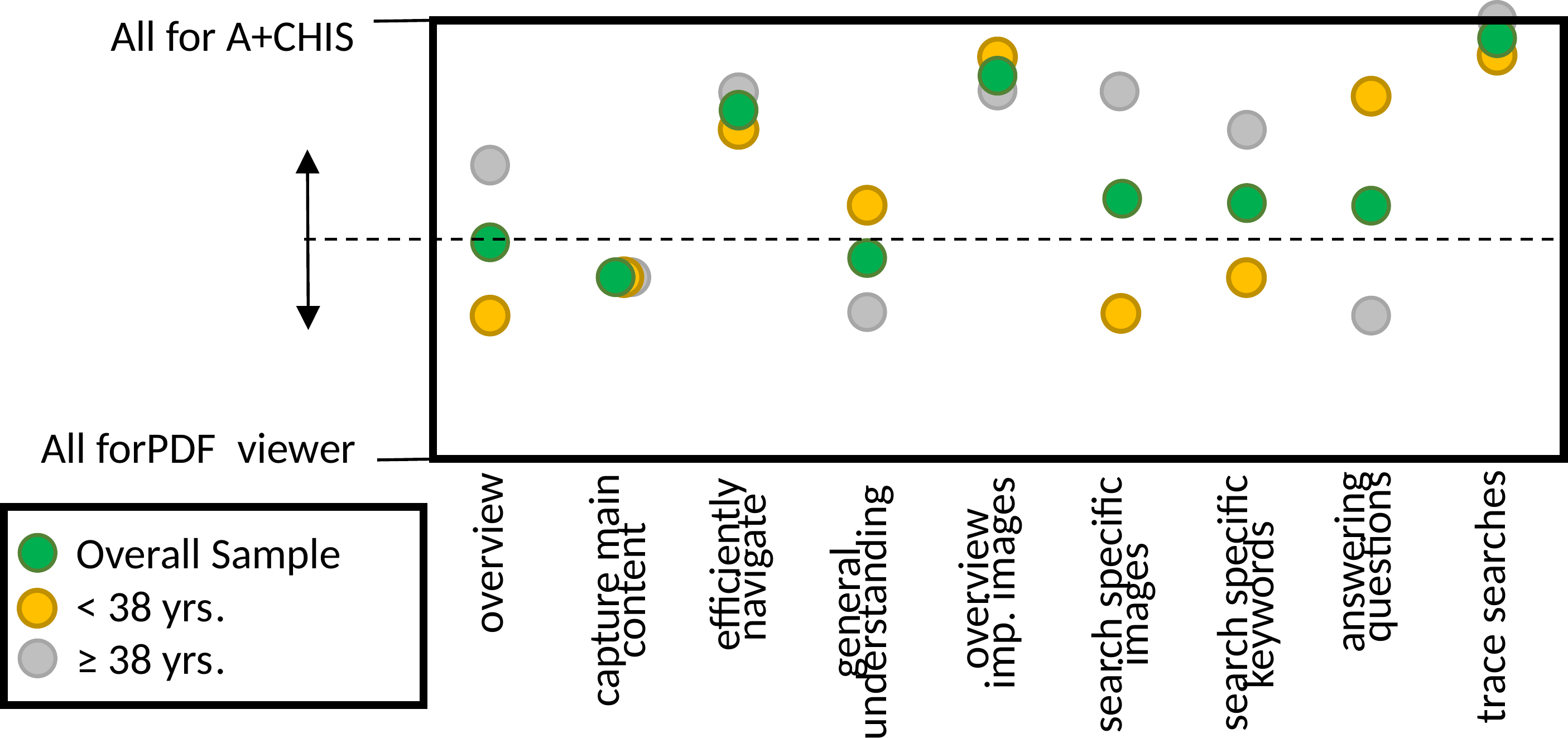}
    \caption{
    The results of the forced choice evaluation between \apluschis\ and the PDF viewer w.r.t. the defined performance goals.
    The performance goals in information processing are sorted along a continuum from more abstract (left) to more specific goals (right). 
    }
    \label{fig:Forced Choice}
\end{figure}

\paragraph*{Performance Goals}
Besides describing the sample as a whole, as part of our explorative research, we compared the results of several subgroups: 
(i) female vs. male participants, and - based on a Median-split - 
(ii) `younger' vs. `older' participants, `higher' vs. `lower' levels of self-assessed 
(iii) prior knowledge of \acrshort{ttwodm}, 
(iv) computer and software skills, and 
(v) experiences with visualizations. 
Only the comparison between the two age-groups shows some differences at face level. 
All other pairs of `sub-groups' are rather similar in their decisions with regards to the \emph{forced choice} items and thus, we do not report the according results.
\figref{fig:Forced Choice} illustrates the findings of the forced choices participants had to make between the PDF viewer and \apluschis\ regarding defined performance goals. 
If participants chose \apluschis\ as their information source of choice for a performance goal, it received a value of `+1'. 
Conversely, `-1' was the value for their choice of the PDF viewer and `0' for an indiscriminate choice. 
Thus, the ordinate ranges from `-12' (all participants chose the PDF viewer) to `12' (all participants chose \apluschis). 
In \figref{fig:Forced Choice}, the values for the whole sample are reported with green circles, while grey circles represent only the older participants ($n = 6$, $\geq 38 \unit{yrs.}$) and orange circles only the younger ones ($n = 6$, $< 38 \unit{yrs.}$).

The results show the potential of \apluschis\ with regard to the fulfillment of the performance goals. 
The values for all nine performance goals are either close to the horizontal `middle-line' (indicating that participants are equally inclined towards \apluschis\ and PDF viewer and/or are indifferent) or clearly above, such as for \emph{efficiently navigating} through different topics of the content, getting an \emph{overview} of the most \emph{important images}, or \emph{tracing searches}. 
We could also observe differences between the two age groups; however, this should not be over-interpreted due to the small sample size but should be given consideration in future research. 
Overall, \apluschis\ is evaluated as a good tool for information searches compared to the PDF viewer, despite participants' familiarity with the latter.

\paragraph*{Non-Linear Content Exploration} 
One area of particular interest was to find out how non-linear content exploration was received by participants. 
Think aloud during the \cwt\ tasks and the subsequent interviews showed that the system arouses curiosity and motivates further exploration of system features and contents. 
The participants perceived the search as enjoyable and found the aesthetic design with colors appealing. 
This applies in particular to the \WordCloud, which manages to cultivate curiosity about further topics, as one participant explained: \emph{‘[...] I can imagine that if there is such a \WordCloud\, I would at least take a look at what topics there are, whether I missed something that would interest me. And I would rather do that than browse through the brochure [...]’ (\PTen)}. 
It is an added value compared to a non-interactive system, that the \WordCloud\ fosters engagement with the content further via making interesting and frequent terms more visible.

However, some participants felt that they could not explore the content effectively because they were unsure about how to use the system due to its novelty. 
In particular, the lack of a familiar linear structure was found confusing and made it difficult for some participants to keep track of the content. 
As one participant noted: \emph{‘[...] I’d like to know what’s in it. 
I don’t know what to expect, what the tool offers me.
It doesn't give an impression of what it actually is now. 
At least not a quick one. 
I don’t have that much patience for it [...] (\PSeven)’}. 
It seems that the open structure overwhelmed some participants, leading to a sense of inadequacy to complete the information search. 
More support through assistance in the \apluschis\ system could help users overcome these initial uncertainties so that they can capitalize on the strengths mentioned above and utilize the advantages to the fullest.

\section{Interaction Analysis} \label{sec:interaction-analysis}
Our data collection efforts, including the on-screen and audio recordings as well as the application of the think-aloud method (\secref{sec:procedure}), yielded valuable insights and enabled us to evaluate the interaction data of the 12 participants during the 11 \acrshort{cwt} tasks in more detail. 
Thus, the analysis of these interaction data is a secondary analysis of the data obtained during the \acrshortpl{cwt} as part of the formative evaluation described in the previous section.

These interactions, such as clicks, scrolls, and key-presses, which are performed in order to derive a new insight, are referred to as \emph{(insight) provenance} \cite{gotz2009characterizing} and are a vital cue for the analysis for cognitive processes (\secref{sec:background}).
After describing the steps involved in capturing our provenance data such as tools and components used by the participants and the processes involved (\secref{sec:tools-and-processes}), the obtained records were analysed on a quantitative basis (\secref{sec:cwt-results}). 
Lastly, we will show how an interactive visual analysis of such provenance data can be enabled through custom-made visual analytics systems (\secref{sec:provenance-visualization}).

\subsection{Background} \label{sec:background}
The underlying working hypothesis of the analysis of interaction data was inspired by behavioral mapping, by process models in the field of information seeking and retrieval (e.g., Joseph et al.~\cite{joseph2013models}), as well by the work of Pohl et al.~\cite{pohl2016using} who applied a lag-sequential analysis (e.g., Bakeman and Gottman~\cite{bakeman1997observing}) to investigate interactions, sequences of interactions and users' activities and processes when engaging with a visualization system. 
Behavioral mapping is a well-established research method where the paths, movements, and activities within a physical space of participants (when carrying out certain tasks) are recorded and transferred onto a map for further analysis. 
As an example, Shepley~\cite{shepley2002predesign} investigated the ways of staff members and their time spent on walking from activity to activity at a neonatal intensive care unit. 
One of the goals of behavioral mapping is to re-arrange the physical space and workstations to avoid unnecessary paths and to make the workflow more efficient. 
The underlying principles, questions, and metrics have been transferred to the virtual space of the \apluschis\ platform and the \acrshort{cwt} tasks, having in mind exploratory research questions such as: \emph{What are the processes and interactions of the participants? 
Are there efficient and inefficient participants with regards to specific tasks, what are their characteristics and how could we support this sub-group of users? 
Could inefficient series of loops and cycles be avoided by a re-arrangement of the tools and components within the platform, by providing more guidance support, or by highlighting central tools and components which are used by a majority of participants across several tasks?} 
To answer such questions as these, quantitatively and by means of metrics from behavioral mapping~\cite{ng2016behavioral} and graph theory (e.g. \emph{centrality}), a formal description of the interaction data of each \cwt\ task per participant, the tasks across all participants, as well as the participants across all tasks are required.

The interaction data of one or more participants or for one or more tasks can be graphically represented as a directed, labeled multi-graph, with the tools as vertices and the processes that lead from one vertex $n$ to a consecutive vertex $n+1$ as edge-labels (or vice versa).

\subsection{Tools and Processes} \label{sec:tools-and-processes}
Before the data processing of the on-screen and audio recordings, a comprehensive list of (cognitive) processes have been defined.
Overall, 28 processes have been pre-defined by the three investigators of the formative evaluation study, whereas two additional processes had to be included during the coding process to ensure that all processes are covered by the raw data. 
Some examples of these processes include \emph{commenting} (when evaluating pictures), \emph{reading} (if text passages have been reproduced verbatim), \emph{interpreting} (concerning pictures and text passages, if participants summarized or evaluated the content in their own words), \emph{pause} (if participants did not do or say anything), but also more basic processes when interacting with the \apluschis\ platform, such as \emph{scrolling}, \emph{sliding} (through the pictures of the \is), or \emph{hovering} and \emph{click on} (e.g. a certain term in the \wc). 
The tools and exploration subsystems (\secref{sec:proposed-design}), have been further distinguished by considering also the chapter of the brochure~\cite{aok} into account; i.e., instead of distinguishing ‘only’ between \ImageSlider, \WordCloud, etc., it has been further differentiated between the preview (small) and the enlarged \ImageSlider\ as well as the \WordCloud, \Tilebar, \Snippets, etc. for each of the seven chapters of the brochure, resulting in 86 ‘tools’.
The coding of the on-screen and audio recordings has been to a large extent carried out individually by the above-mentioned investigators.
However, in cases the investigators were not absolutely sure on how to interpret a participants activity and to code it into one of the pre-defined processes, he or she noted the time-stamp and the coding of such ambiguous activities has been done collaboratively by reaching a consensus.
The processes have been coded only if they exceeded a duration of around $1 \unit{s}$. For the computation of several metrics and further analysis, the individual interaction data for each of the tasks has been represented as sequence of 
$\langle \mathit{tool}_{\mathit{src}}, \mathit{process}, \mathit{tool}_{\mathit{tar}} \rangle$
triples.

\subsection{Results and Discussion} \label{sec:cwt-results}
Overall, i.e., across all 12 participants and 11 \cwt\ tasks, 1,870 processes (=edges) have been observed, whereas 21 out of the 30 processes were applied at least once by the participants. The three most frequently applied processes are \emph{scrolling} ($n = 268$), \emph{sliding} ($n = 240$) and \emph{click on} ($n = 235$). 
On average, a process took $4.18 \unit{s}$ (\emph{Mdn} = $2 \unit{s}$, \emph{SD} = $5.03 \unit{s}$), ranging from $1$ to $64 \unit{s}$. 
The tasks (\tableref{tab:CWT}) where the most processes have been applied by all participants are \emph{IS2} ($n = 337$), \emph{WC2} ($n = 267$) and \emph{IS3} ($n = 242$); whereas the tasks with the least amount of processes are \emph{WC4} ($n = 65$), \emph{TiB1} ($n = 77$) and \emph{WC1} ($n = 94$). 

With regards to the `tools', 75 out of the 86 differentiated tools have been applied at least once by the participants, whereas the three most frequently applied tools are the \emph{\isL}\ (the \ImageSlider\ in enlarged form) \emph{for chapter 6} ($n = 308$), the \emph{\snps\ for chapter 1} ($n = 124$) and the \emph{\Searchbar} ($n = 103$). 
To put these numbers in context, the $75$ tools which have been applied at least once, were visited $2,002$ times across all participants and \acrshort{cwt} tasks. 
The three most applied tools, i.e., those with the highest \emph{centrality}, cover around a quarter of all incoming and outgoing edges.

The triple sequence $T$ allows to easily evaluate further metrics, such as the number of
(i) \emph{loops}, i.e., $|\{\langle s, ., t \rangle \in T : s = t \}|$,
(ii) \emph{multiple edges}, i.e., $|\{\langle s, ., t \rangle \in T : (\exists \langle s', ., t' \rangle \in T\setminus \{\langle s, ., t \rangle\})[s = s' \wedge t = t']\}|$,
and (iii) \emph{identical triples}, i.e., 
$|\{\mathbf{t} \in T : (\exists \mathbf{t}' \in T\setminus \{\mathbf{t}\})[\mathbf{t} = \mathbf{t}']\}|$.
Overall, $1,870$ triples have been identified. 
$1,030$ of them are \emph{loops}, comprising $59$ (of the $75$) tools. 
The three tools with the most \emph{loops} are the \emph{\isL\ for chapter 6} ($n = 288$), the \emph{\snps\ for chapter 1} ($n = 91$) and the \emph{\snps\ for chapter 3}. 
These three cases are identical to the three most prominent \emph{multiple edges}. $1,479$ triples represent \emph{multiple edges} with $197$ unique tool combinations.
Finally, $1,177$ triples represent \emph{identical triples}, with $171$ distinct instances. 
The three most prominent ones are subsets of the above mentioned \emph{loops} and \emph{multiple edges} concerning the tool \emph{\isL\ for chapter 6}, with the processes \emph{sliding} ($n = 170$), \emph{commenting} ($n = 53$) and \emph{viewing} ($n = 41$).

The fact that certain chapters of the brochure occur several times within the top-three of the applied tools, the \emph{loops}, \emph{multiple edges} and \emph{identical triples}, is of course caused by the concrete \cwt\ tasks (e.g., the tasks \taskTibThree\ and \taskIsTwo\ which specifically ask for a comparison between the content in chapter 6 and other chapters; see \tableref{tab:CWT}).

\begin{figure}[ht!]
    \centering
    \includegraphics[width=\mediumwidth\linewidth]{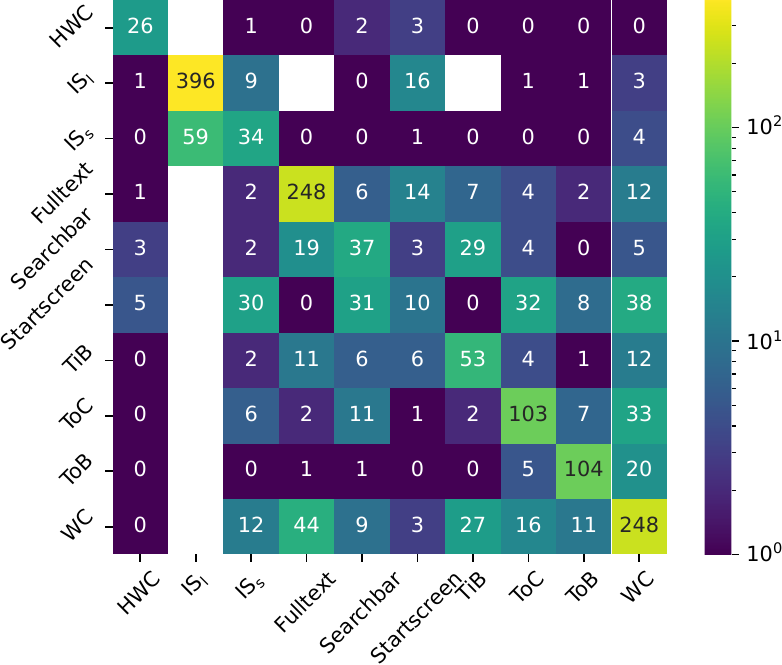}
    \caption{
    The adjacency matrix of transitions between the `high-level' tools. 
    The entries in the main diagonal corresponds to our notion of \emph{loops}, while \emph{multiple edges} populate the remaining cells.
    The cells reflecting transitions which are technically not possible are intentionally left blank.
    }
    \label{fig:adjacency-matrix}
\end{figure}

The transitions between tools (agnostic with regards to chapters) are shown in the matrix in \figref{fig:adjacency-matrix}. 
The \emph{loops} correspond to the main diagonal, while all cells with a value $> 1$ contain \emph{multiple edges}. 
Please note that several transitions from one tool to another are technically not possible; in particular, several transitions from or to the \isL. 
When distinguishing only between the more broadly defined $10$ tools as in \figref{fig:adjacency-matrix}, out of the $1,870$ triples, $1,259$ are \emph{loops}, whereas all tools are affected, $1,802$ triples represent \emph{multiple edges}, with $58$ distinct tool combinations, and $1,616$ represent \emph{identical triples}, with $221$ unique instances. 
The three most observed \emph{identical triples} are 
$\langle \mathit{\isL}, \mathit{sliding}, \mathit{\isL} \rangle$ ($n = 221$), 
$\langle \mathit{\wc}, \mathit{scrolling}, \mathit{\wc} \rangle$ ($n = 77$), and 
$\langle \mathit{\wc}, \mathit{scanning}, \mathit{\wc} \rangle$ ($n = 74$). 

\mediumwidth

The often observed \emph{loops}, \emph{multiple edges}, and \emph{identical triples} with regards to the 
\emph{\isL}, reveal some potential for improvement to make the information search for users more efficient. 
One potential improvement approach has been suggested by a participant in the course of the formative evaluation study (\emph{\POne}: An \emph{Image Tile Display} that allows for more efficient scanning through the images of a certain chapter at once).
%

\subsection{Provenance Visualization} \label{sec:provenance-visualization}

While a `global' evaluation of the provenance -- e.g., obtaining quantitative measures for \emph{loops} or usage of individual components -- can be conducted based on the raw interaction transcripts, a more in-depth analysis requires dedicated visualizations.
Those would allow the analysis of interactions on a per-user and/or per-task basis and could answer additional questions, such as
`Are the different groups of users observable?' (e.g., depth-first search vs. breath-first search exploration process), `Are there any outliers?' (users whose exploration process differs significantly from all others), etc.
The data's underlying graph structure invites the application of different established visualizations. 
We implement two customized visual analytics tools, a graph as well as a matrix layout, which allow us to investigate different orthogonal aspects of the interaction data.

\begin{figure}[ht!]
    \centering
    \begin{minipage}{\mediumwidth\linewidth}
        \includegraphics[width=0.49\linewidth]{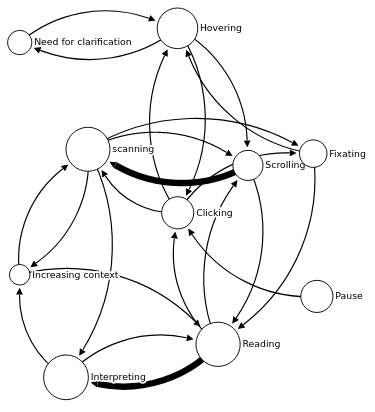} 
        \includegraphics[width=0.49\linewidth]{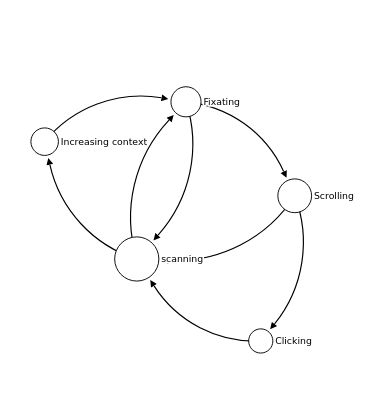} \\
        \includegraphics[width=\linewidth]{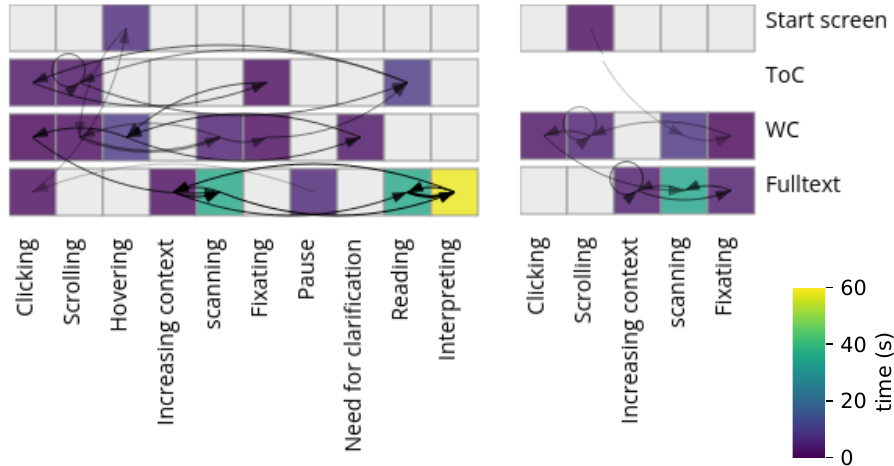}
    \end{minipage}
    \caption{
    Provenance visualizations for two very different users, \POne\ (left) and \PTwelve\ (right), working on task \taskWcThree. 
    Their respective provenance graphs (top) reveal the alternations between processes, while the provenance matrices (bottom) clearly show the alternations between levels of visual granularity.
    }
    \label{fig:provenance-vis}
\end{figure}

\paragraph*{Provenance Graph}
In a \emph{Provenance Graph}, we visualize a user's alteration and switching between different processes; i.e., in a directed weighted graph, we show how often a user switched between different processes (\figref{fig:provenance-vis}, bottom). 
With this type of visualization, it is possible to, for instance, spot if a user goes back and forth between two different processes or if they go through the same sequence of processes over and over (cycles). 
We indicate the time spent on specific processes through the size of the respective node and, conversely, the number of transitions between processes though the thickness of the respective links.

\paragraph*{Provenance Matrix}
Besides the graph representation, we also visualize the provenance information in a matrix layout where the sorted rows (\Startscreen/\dl\ $>$ \toc\ $>$ \wc/\hwc/\is\ $>$ \tob/\tib\ $>$ \snps/\fullt) represent the high-level tools at  different levels of visual granularity and abstraction (overview to closeup $\sim$ top to bottom). 
The columns reflect the different processes, sorted by their type (from basal/technical processes such as `scrolling' to cognitive/psychological processes such as `interpreting').
The matrix cells exhibit a color-coding, indicating the overall time a user spent on a tool-process pair.
Additionally, we show the sequence of the exploration with arrows spanning consecutively visited cells. 
As the display of `all' arrows would overload the visualization, we display only the most recent transitions, with the recentness modelled by an alpha-drop-off. 
Hovering over a cell triggers a tooltip which lists all the interaction triples responsible for said cell; i.e., as opposed to the provenance graph, the matrix is able to reveal correlations between processes and components, i.e., it shows 
(i) at which levels of visual abstraction a user predominantly operates, 
(ii) which processes they carry out at which level, 
(iii) which processes they carry out using which component, and 
(iv) whether they exhibit a rather vertical ($\sim$ depth-first search) or horizontal ($\sim$ breath-first search) exploration behavior.

\paragraph*{Use Case Examples}
Finally, we investigate the effectiveness of the proposed provenance visualizations for the visual analysis of tasks conducted by different users.
To this end, we take a look at one exemplar task (\taskWcThree, Version 1: `How does type I diabetes mellitus develop?') which should both, encourage an `open' exploration process, and result in comparable interactions amongst users. 
We compare the respective provenance visualizations of two participants (\POne\ and \PTwelve) with vastly different visual analytics 
proficiencies. 
It took \POne\ $188 \unit{s}$ to successfully complete the task, while \PTwelve\ required only $63 \unit{s}$.
\figref{fig:provenance-vis} shows the respective provenance graph and matrix, illustrating the interactions captured while working on said task. 
Even at a first glance it is obvious that \POne\ underwent a much more laborious exploration as \PTwelve\ which is in line with our impression during the supervised evaluation.

On a closer look, we can see \PTwelve\ used only 5 distinct processes without much back-and-forth between any of them. 
An inspection of the respective arrows in the provenance matrix also reveals that this participant generally moved from basal processes at a high level of abstraction (top left corner of the matrix) to rather cognitive processes at detail level (bottom right corner of the matrix). 
This is an expected exploration pattern of a competent information seeker.

\POne, on the other hand, required not just more time overall, but had a lot more back-and-forth between different technical and cognitive processes, i.e., `Scrolling' $\rightleftarrows$ `Scanning' and `Reading' $\rightleftarrows$ `Interpreting'. 
This interaction pattern is also confirmed by the provenance matrix which further reveals that \POne\ has several movements from low- to high abstraction levels such as from \fullt\ to \wc, or from \wc\ to \toc. 
We take these bottom-to-top patterns as a cue for an individual who ran into an exploratory wrong track, hence they had to backtrack to a higher level in order to find the correct path to their desired information.

The next research focus will be on the automatic analysis and the clustering of users based on their interaction patterns, as this information can be leveraged to propose adequate visualization to them \cite{gotz_behavior-driven_2009}.

\section{Overall Discussion} \label{sec:discussion} 

Our exploration system allows users to navigate through documents and adapt the visual representation and level of detail by using well-known visual analysis techniques such as word clouds, topic models, tile bars, and keyword search. 
The system provides the users with a two-fold document exploration: a traditional linear and non-linear document navigation by switching between content and detail. 
Furthermore, it allows users to follow both the edited content of a given document (supervised structure) as well as an automatically computed topic models (unsupervised structure). 
To the best of our knowledge, there are few empirical studies on the cognitive and motivational aspects of using document visualizations (\eg\ tile bars and word clouds, with those of linear document readers). 
Our evaluation is a first confirmation that our approach could foster interest and heighten curiosity by using a distant-reading approach for exploring the content of interest more efficiently.
Further key findings are that the participants in general enjoyed the non-linear content exploration, even if some participants would have needed more support functions at least in beginning of use. 
There were mixed results regarding intuitiveness of the \WordCloud\, whereas the \Tilebar\ and the \ImageSlider\ have been evaluated as being intuitive; which is also reflected in their agreement to the question if they would use these components again.

\begin{figure}
    \centering
    \includegraphics[width=0.6\linewidth]{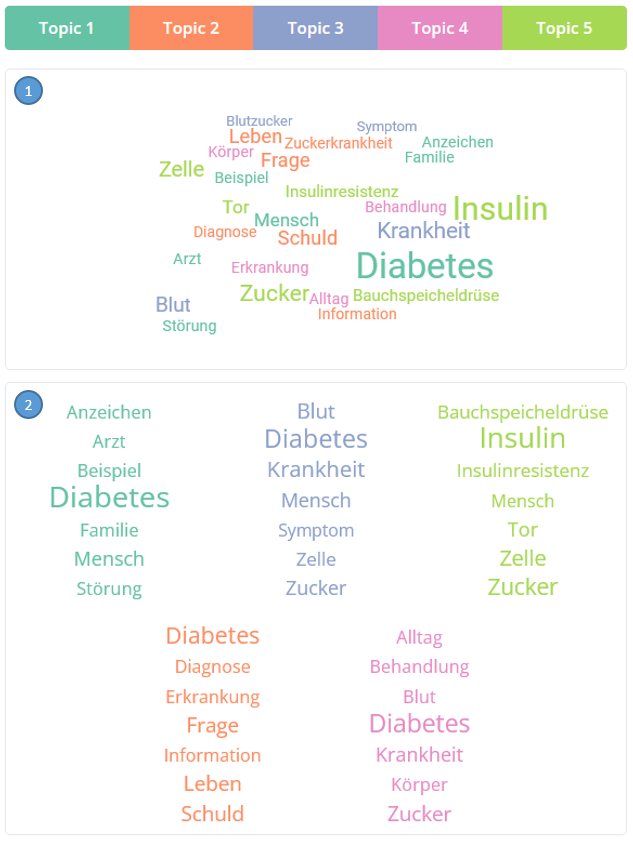}
    \caption{Adaptive word cloud layouts: for improving topic understanding the arrangement can be switched from Wordle layout (1) to an alternative list representation (2).}
    \label{fig:wc-tc}
\end{figure}

Our study showed that users did not make great use of the topic model structure. 
This may be partly due to the unfamiliar representation of topic models. 
Recently, some studies have investigated the impact of word clouds for topic understanding \cite{10.1162/tacl_a_00042} and keyword summaries \cite{8017641}. 
Word clouds, as it transpired, are particularly useful for quickly identifying the most common and frequent terms, while disadvantages may arise in decoding numeric values from font sizes for larger sets of keywords. 
An alternative visualization could be a simple word list with a frequency encoding (\eg\ font size, bars) that represent each topic. 
\figref{fig:wc-tc} shows such an alternative representation (2) to improve topic understanding where the keywords for each topic are displayed among each other. 
By using such a layout, overlapping term composition may be included to increase topic understanding and cause confusion as with the previous layout (1). 
Advanced topic model visualization will be considered in future work.

One key element of our system is its ``user tracking'' during document exploration.
In particular, we track which keywords have been explored, with which visual component, and for how long. 
This is considered \emph{important information provenance data} which, in our system, is used for provenance visualizations (\eg\ history word cloud, interaction matrix and graph). 
The latter is an important functionality for content recommendation, and forms the basis for the mitigation of cognitive biases and potentially harmful and wrong pre-conceptions.
The provenance visualizations may reveal emerging difficulties of users during information seeking tasks by showing outstanding patterns of user behavior, \eg\ horizontal, vertical or loop-like patterns. Our evaluation is a first step in this direction. 
To investigate user behavior during the CWT tasks, we integrated the CWT tasks in our provenance visualizations.

\subsection{Future Work} \label{sec:future-work}
The advantages of aggregated document representations warrant further examination in future research; in particular, the specific benefits that can be derived from using aggregated document representations. 
Furthermore, our future work will involve exploring potential misunderstandings that may arise from the highly aggregated nature of certain content presentations.

As evident from \secref{sec:interaction-analysis}, a vital cue which we plan to leverage for adaptive visualizations are \emph{user interactions}.
Research has demonstrated that user interactions can effectively be utilized for recommending particular types of visualizations \cite{gotz_behavior-driven_2009} and even help mitigate cognitive biases \cite{gotz_adaptive_2016}.
An investigation of the user interactions obtained from the \acrshort{cwt} (\secref{sec:cwt-results}) with customized visual analysis tools (\secref{sec:provenance-visualization}) revealed that meaningful interaction patterns, reflecting different types of users, can be observed.
We assume that users can be clustered into cohesive groups using said patterns, which, in turn, reflect a user group's need for specific types of visualizations.
Therefore, an open challenge is the robust and continuous tracking and classification of these interactions.
While this was done manually for the 12 participants of the \acrshort{cwt}, a purely automatic solution is necessary for the long-term.
Our prototype already comprises a tracking of the components a user is interacting with. 
To this end, we leverage the mouse pointer position together with the established assumption that a user's cursor movements are correlated with their gaze~\cite{reichle_ez_2006, buscher_eye_2008}.
The other aspect of our interaction logs -- the process a user is occupied with (\secref{sec:tools-and-processes}) -- is significantly harder to track automatically. 
Even though purely technical processes such as `scrolling' or `clicking' are trivial to track, the capturing of cognitive processes such as `reading' or `interpreting' pose a non-trivial challenge.
Yet, even such can be determined using low-level mouse interactions together with well-defined heuristics~\cite{10.1145/2207676.2208591, kirsh_virtual_2022}.

Additionally, we have planned further user studies. 
These will focus on the planned enhancements of the system, such as the representation of behavioral patterns, which can support users in reflecting on their information-seeking behavior and, thus, potentially contribute to the detection and prevention of biases, especially confirmation bias. 
This will include investigating how different ways of presenting the interactions can promote unbiased information-seeking by also taking into account differences between users, such as different visualization preferences, different states of knowledge or, as already mentioned above, age groups. 
Moreover, the need for support, as found in our current study for some users due to the novelty of the system, will be addressed.

In future work, we also intend to research automatic recommendation and develop adaptation methods based on the current system. 
Such a further developed system version will also be subjected to an extensive case study in order to evaluate the system in detail.

\subsection{A Model for Adaptation of Content and Presentation}

Our initial motivation was to provide an interactive and adaptive \chis. 
These systems should adapt the content and presentation to the users' information needs and preferences. 
Our document exploration design has a number of variables which could be controlled and adjusted by the system to adapt itself to the users' information need and preferences, and recommend views and content. 
To that end, we define a model based on the following dimensions along which automatic adaptation can be performed. 
In future work, we will focus on the prediction of dimensions to use and how to set them for specific users.

\paragraph*{Content Navigation} 
Here, \emph{what to show} when the user is doing a mouse-over on a term of the Word Cloud is to be decided. 
The principal options include (a) go to the best matching full text position (full detail), (b) show the text snippets of multiple matches (an intermediate detail level), or (c) show the tile bar (lowest level of detail) from which the user may pick a finding location. 
The system could determine the answer based on the previous selections made by the user, presuming the user has a stable preference. 
On the other hand, the system could track which background knowledge is already available, and show the higher levels of details for topics which are not well-known.

\paragraph*{Document Structure} 
Our \apluschis\ design shows the section Word Clouds together with the (a) section headings or (b) topic models computed for each section. 
These represent both supervised and unsupervised content structures. 
When the user requests a section, the system might adapt to show either one of these. 
To this end, the system might predict whether the user prefers the traditional, edited document structure given by the headings, or the computed content structure from a topic model. 
The latter provides an opportunity to compute the topic models such as to represent the users' interest and background information.

\paragraph*{Configuration of the History Cloud} 
The system may present to the user either a History Cloud of (a) what has already been explored, or (b) what has not been explored yet. 
The system could predict if the user would want to deepen their knowledge on a particular topic or broaden their horizon into new topics. Depending on the user's intent, the system can choose from which terms to create the history Word Cloud. 
This might also be a good starting point to mitigate biases once they are detected in the users.
It is important to recognize that these are concepts, and thus require specific implementations to monitor, track, and characterize the users' background knowledge, preferences, and interests. 
In order to explore potential implementations and solutions to realize these concepts, both direct and indirect methods will be examined in our future research work.

\section{Conclusion} \label{sec:conclusion}
Currently, a significant number of existing CHIS fall short when it comes to presenting health information in interactive, adaptive, and/or personalized ways. 
In this regard, interactive document visualization techniques are more conducive to enhancing user engagement and promoting a deeper understanding of complex information, thus providing an amplitude of opportunities for improved support of information-seeking tasks. 
We presented a novel and innovative document exploration system that allows users to visually explore health documents at various levels of detail and abstraction. 
We incorporated well-known document visualization techniques, such as \WordCloud s and \Tilebar s, into our design and applied it in the domain of \acrshort{ttwodm}. 
We evaluated our implemented system by performing a formative study based on a \cwt\ and compared our approach with linear and close reading. 
The evaluation results are promising and constructive, and demonstrate that our approach is easy to use and helps in content exploration, further facilitating more interactive content navigation as well as motivating users to engage with our implemented system at different levels of visual granularity and detail. 
We also presented concepts for possible visual adaptation by collecting user interaction data and visualizing this data in two provenance visualizations to reveal specific information needs as well as reading preferences.

\section*{Acknowledgments}
\paperAcknowledgement

\printbibliography

\end{document}